\documentstyle[aps,psfig]{revtex}

\begin{document}
\draft
\title{Eden growth model for aggregation of charged particles}
\author{Y.V. Ivanenko$^1$, N.I. Lebovka$^{1,2}$\thanks{%
lebovka@roller.ukma.kiev.ua}, N.V. Vygornitskii$^1$}
\address{$^1$Institute of Biocolloidal Chemistry, NAS of Ukraine, 42, blvr.\\
Vernadskogo, Kyiv, 252142, Ukraine\\
$^2$Kyiv Mogyla Academy University, 2, vul. Scovorody, Kyiv, 252145, Ukraine
\\
Received
\today }
\maketitle

\begin{abstract}
The stochastic Eden model of charged particles aggregation in
two-dimensional systems is presented. This model is governed by two
parameters: screening length of electrostatic interaction, $\lambda $, and
short range attraction energy, $E$. Different patterns of finite and
infinite aggregates are observed. They are of following types of
morphologies: linear or linear with bending, warm-like, DBM (dense-branching
morphology), DBM with nucleus, and compact Eden-like. The transition between
the different modes of growth is studied and phase diagram of the growth
structures is obtained in $\lambda, E$ co-ordinates. The detailed
aggregate structure analysis, including analysis of their fractal
properties, is presented. The scheme of the internal inhomogeneous structure
of aggregates is proposed.
\end{abstract}




\section{Introduction}

\label{S1}There are a lot of computer models aimed at simulation of
the natural patterns occurring in colloidal aggregation,
electrochemical deposition, dielectric breakdown, bacterial colony
growth, viscous fingering, spreading, etc.\cite{family}. Such models
allow to explore the process of pattern formation in real physical
systems and they are based mostly on the Eden model \cite{eden} and
the model of diffusion limited aggregation (DLA) \cite{witten}, with
due account of particles drift, convection flow, diffusion, external
field influence and specific interactions between the particles. The
random successive growth model \cite {wu}, two-scale drift-diffusion
model \cite{john}, diffusion-migration \cite {huang} and
convection-migration models \cite{elebacher} are good examples of such
models. These models describe the most important pattern morphologies
observed in various non-equilibrium systems, such as DLA-like,
dendritic, needle, treelike, dense-branching, compact, stingy, spiral
and chiral structures \cite{nagatani}.

The relationship between the structure and the growing pattern
morphologies, growth mechanisms and peculiarities of interaction
between the elements of growing structure seems to be the main problem
in this field. The local interaction is introduced into the screened
growth model \cite{rikvold}, in which the growth probability is
determined by multiplicative screening contributions of the occupied
sites nearest to the site of growth. In this model, the fractal
dimensionality of the cluster $D_f$ is determined by the form of the
screening function and may be varied in the wide range. Ausloos et al.
have studied the magnetic Eden model \cite{ausloos} and magnetic
diffusion-limited aggregation model \cite{vandewalle} for the
Izing-like nearest-neighbour interaction between the spins. Each of
these models leads to the formation of different types of aggregates
for different values of parameters controlling the growth processes.
However, inclusion of the long-range particle interactions, for
example, of dipolar type, requires considerable CPU-resources. At
present, such simulations are done only for the systems of moderate
size \cite{mors}. However, this question is very important for the
interpretation of the experimental data on magnetic particles
aggregation \cite{helgessen}.

The study of the models of charged particles aggregation is also of a
great interest. It is especially true in relation with the study of
particle surface charge effects on the mechanism of colloidal
aggregation \cite {fernandez}, space charge effects in electrochemical
deposition experiments \cite{chazalviel}, pattern formation mechanism
in discharge systems \cite {breazeal} and possible fractal structure
of ball lightning \cite{smirnov1}. The presence of the long-range
interactions can produce the qualitatively new effects, not usual for
the systems with the short-range interactions. Thus, systems with the
spatial charge distribution are characterised by the limited
stability. It is well known, that the electrostatic interactions cause
the loss of stability of charged droplet \cite{landau}, and may cause
instabilities of charged polymer systems \cite{kantor}, aerosols
\cite{reist} , etc. For the non-equilibrium growth condition, one may
expect the occurrence of such phenomena as large-scale fluctuations,
lacunarity or growth of finite size aggregates.

In this work, we propose the stochastic Eden-like lattice model of charged
particles aggregation. Particles overcome the electrostatic repulsive
barrier created by the aggregate and stick to it due to existence of a
short-range attraction force. We observe the formation of aggregates with
different morphologies, such as: linear or near-linear, linear with bending,
warm-like, DBM or DBM with nucleus, and compact Eden-like.The transition
between the different modes of growth is governed by the two model
parameters screening length, $\lambda $, and attraction energy, $E$.

The structure of the paper is as follows. In Section \ref{S2} we describe
the model. Section \ref{S3} reports the results of simulation and their
discussion, including examples of the clusters, phase diagram, analysis of
the radial cluster structures and their fractal properties. Section \ref{S4}
contains our conclusions. Appendix \ref{S6} presents some simple
electrostatic estimations for the model.

\section{Model}

\label{S2}The model is based on the standard two-dimensional stochastic Eden
model \cite{eden}. We assume that all the particles bear the same electric
charge $q$. A particle located at point j in the proximity of the aggregate
interacts with all the particles of aggregate through the long-range Coulomb
repulsion, $U$
\mbox{$>$}
0, and short-range 'glue' attraction, $E$. We also take into account the
possibility of screening of electrostatic repulsion, and introduce the
screening length $\lambda $, in order to switch off the influence of the
aggregate particles located further then at the distance $\lambda $ from
point $j$. The energy of electrostatic repulsion may be defined as:

\begin{equation}
U(j)=\sum\limits_iq^2a_{ij}/r_{ij},  \label{e1}
\end{equation}

where $r_{ij}$ is a distance between the $j$-point and $i$-particle of the
aggregate, $aij=1$ for $r_{ij}<\lambda $ and $a_{ij}=0$ in any other cases.
The cluster growth takes place on the simple square lattice. The lattice
step and the particle charge are assumed to be equal to $1$. So, in this
model, the cluster growth is governed by two parameters: the screening
length, $\lambda $, and the attraction energy, $E$ and we may consider the
values of $\lambda ,U$ and $E$ as dimensionless.

A cluster grows according to the following scheme:

a) the seed particle of the cluster is placed at the lattice point $x=y=0$;

b) the cluster perimeter site, $j$, is chosen randomly (the perimeter of the
cluster is a set of unoccupied lattice sites nearest to the cluster); the
value of $U$ for the $j$-site is calculated and compared with the $E$-value;

c) if $U\leq E$, then the $j$-site becomes the site of a cluster, and
determination of a new cluster perimeter site takes place; otherwise, if
$U>E $, the $j$-site is declared as unoccupied forever;

d) then the step b) is reiterated.

These steps run until the cluster reaches either the lattice size, or the
desirable size, or until the cluster growth termination. The last case may
be observed when the following condition becomes true for all the sites of
cluster perimeter:

\begin{equation}
U>E.  \label{e2}
\end{equation}

The lattice size varied from $400$x$400$ to $5000$x$5000$, the maximal
number of particles in the cluster did not exceed $3^{.}10^5$. The results
were averaged over the $10\div 30$ different initialisations.

\section{Results and discussions}

\label{S3}

\subsection{Phase diagram}

\label{S31}Fig. \ref{fig1} presents examples of clusters that appear at
different values of the screening length, $\lambda $, and attraction energy, 
$E$. All patterns were obtained on the basis of cluster images on $400$x$400$
lattice. Small circle in the centre of each pattern denotes the seed
particle. The plane of $\lambda $ and $E$ values is approximately divided
into several regions that correspond to the observed cluster types. These
regions are schematically presented on the phase diagram in Fig. \ref{fig2}.
The numbers of patterns of the Fig. \ref{fig1} correspond to the numbers in
black squares put on the phase diagram. We discern the following regions of
the phase diagram:

${\bf L}_{{\bf f}}$ is the region of linear or near-linear finite cluster
growth;

${\bf Lb}_\infty $ is the region the infinite linear clusters or linear
clusters with bending;

${\bf W}_{{\bf f}}$ is the region of finite warm-like clusters; this type of
structure is represented by the patterns 1, 5, 9, 13, 17 of Fig. \ref{fig1}
and similar structures are reported to occur for the random successive
growth model \cite{wu};

${\bf W}_\infty $ is the region of infinite warm-like clusters; this
structure type examples are presented by the patterns 2, 6, 10, 14 of Fig. 
\ref{fig1};

${\bf DB}_\infty $ is the region of lacunary clusters, clusters with
developed branch structure and clusters of DBM-type \cite{nagatani}
structures growth; in this region, clusters may possess a central dense part
or a nucleus (see, patterns 3, 4, 7, 8, 11, 12, 15, 16, 18, 19, 20 of Fig. 
\ref{fig1});

${\bf C}_\infty $ is the region of compact clusters or compact clusters with
cavities growth; in the limit $E\rightarrow \infty $ these clusters are the
clusters of standard Eden model.

The case of $\lambda =\infty $ (see 21-24, Fig. \ref{fig1}) is shown above
the main diagram (Fig. \ref{fig2}). There are no regions of infinite cluster
growth and no compact cluster growth for $\lambda =\infty $ (except for the
limiting case of $E=\infty $, when the growth of Eden-like clusters takes
place).

Thus, on increasing of $E$ at fixed $\lambda $ value, we observe the
following sequences of regions

\begin{equation}
{\bf L}_{{\bf f}}{\bf \rightarrow Lb}_\infty
{\bf \rightarrow Wf\rightarrow W}_\infty
{\bf \rightarrow DB}_\infty {\bf \rightarrow C}_\infty  \label{e3a}
\end{equation}

when $\lambda $ is finite, and

\begin{equation}
{\bf L}_{{\bf f}}{\bf \rightarrow Lb}_{{\bf f}}
{\bf \rightarrow W}_{{\bf f}}
{\bf \rightarrow DB}_{{\bf f}}  \label{e3b}
\end{equation}

when $\lambda $=$\infty $.

The main distinctive feature of the phase diagram in Fig. \ref{fig2} is the
existence of regions with {\it finite} and {\it infinite} cluster growth. It
appears merely due to the long-range character of Coulomb interactions. When
a cluster grows, its total charge increases and it results in increase of
the electrostatic repulsion energy at the points of new particle attachment,
which are located on the cluster perimeter. In the certain cases, when the
cluster sufficiently large grows, the condition of cluster growth
termination (\ref{e2}) realises for all the points of cluster perimeter. We
characterise the maximal size of a cluster by its maximal radius
$R_{\max }$, which is simply the distance from the central seed
particle to the most remote particle of the cluster. The full
manifestation of the long-range character of Coulomb interactions
occurs only in the cases when the value of $\lambda $ exceeds the
spatial dimension of the cluster, i.e. when $\lambda \geq R_{\max }$.
This condition is always true when $\lambda =\infty $, which is the
cause of only finite cluster growth.

How it is possible to discern the different regions of the phase diagram
presented in Fig. \ref{fig2}? Some conclusions about localisation of
different regions can be made on the base of the visual analysis of patterns
similar to that presented on Fig. \ref{fig1}. For fixed $\lambda $, the
increase of $E$ leads to the growth of different type clusters. Thus, in
${\bf L}_{{\bf f}}$ region, the length of cluster increases with $E$
and in ${\bf L_b}$ region the number of cluster bendings increases
and the length of the linear cluster part decreases with the increase
of $E$. However, the visual analysis does not allow us to make an
exact estimation of region localisations. This method is effective
only for determination of the sharp boundary between the region ${\bf
DB}_\infty $ and region ${\bf C}_\infty $. Note that this boundary can
also be localised by means of simple electrostatic estimations from
which we obtained for the transition energy $ E_c$ between region
${\bf DB}_\infty $ and region ${\bf C}_\infty $ the following
relation: $E_c=\pi \lambda $ (See Appendix, Eq. (\ref{eA5})).

The boundaries of ${\bf L}_{{\bf f}}$, ${\bf Lb}_{{\bf \infty }}$, ${\bf W}_{
{\bf f}}$ and ${\bf W}_{{\bf \infty }}$ regions are not defined clearly and
the transition from one phase to another is not sharp and appears to be
rather smeared. Due to the finite-size effects near the boundaries of
regions, we observe strong fluctuations of cluster sizes and morphologies.
The width of such transition zones between the different regions may vary
and is difficult to determine, however, in our simulations it does not
exceed $\Delta E\approx 0.2$ for fixed $\lambda $. On moving from region
${\bf L}_{{\bf f}}$ to region ${\bf L}_\infty $ at fixed $\lambda $ value,
the closer we are to region ${\bf L}_\infty $, the higher is the probability
of an infinite linear cluster growth. Correspondingly, the probability of
finite linear cluster growth is lower. For estimation purposes, the boundary
line may be defined as a line of approximately equal probabilities of
infinite and finite clusters formation. This is true also for the cases of
${\bf Lb}_\infty {\bf -W}_{{\bf f}}$ and ${\bf W}_{{\bf f}}{\bf -W}_\infty $
boundaries. Unfortunately, such approach is not acceptable for boundary
localisation between ${\bf W}_\infty $ and ${\bf DB}_\infty $ regions.

To estimate the region boundaries more precisely, we have carried out the
configuration analysis of cluster structures and studied the behaviour of
the maximal cluster radius $R_{_{\max }}$.

\subsection{Configuration analysis and maximal cluster radius}

\label{S32}Transitions between the different regions of phase diagram may be
also analysed with the help of configuration analysis. The idea is to find
some correlation between the pattern type and the probability of certain
nearest neighbour configuration appearance in that pattern. Any cluster on a
simple square lattice consists of the particles that have 1, 2, 3 and 4
nearest neighbours. For the purposes of configuration analysis, we have
discerned four types of configuration presented in Fig. \ref{fig3}.

It is easy to see that the configurations of type (1) correspond to the
points of cluster growth termination, the linear clusters consist mostly of
type (2) linear configurations, the clusters with bending and warm-like
clusters are the mixtures of type (2) and type (3) configurations of
particles and the particles of compact and branched clusters form mostly the
configurations of type (4).

The fractions of particles in ($i$) type
configurations, $P_i$, were calculated as the ratio of the number of
particles having one neighbour in configuration (1), two neighbours in
configuration (2), two neighbours in configuration (3) and three or four
neighbours in configuration (4) respectively, to the total number of cluster
particles.

To localise more precisely the boundary between the above mentioned finite
and infinite cluster regions, we have also studied the dependency of maximal
cluster radius $R_{_{_{\max }}}$versus attraction energy $E$ for fixed
values of screening length $\lambda $. The results were averaged over
$10\div 30$ different initialisations for each chosen pair of values $\lambda
${, }${E}$ and the maximal lattice size used this work was $5000$x$5000$. A
cluster reaching the lattice boundaries was considered as infinite. Such
clusters were observed only for $\lambda <\infty $, while, as it is
mentioned above, for $\lambda =\infty $ all clusters are only finite, they
stop to grow far beyond the limits of our system.

Figures \ref{fig4}-\ref{fig5} show the fractions of particles with relevant
nearest neighbour configurations, $P_i$ (a) and the maximal cluster radius,
$R_{\max }$ versus $E$ (b) for $\lambda =30$, $\infty $, respectively. These
figures correspond to the continuus move along the $E$ axis across the phase
diagram presented at Fig. \ref{fig2} at fixed values of $\lambda $.

Note that at finite $\lambda $ the transition between the finite and
infinite growth regions in the sequence (\ref{e3a}) is obvious at $R_{_{\max
}}$ versus $E$ curve (Fig. \ref{fig4}b). This is the way to fix the
transition boundaries between the ${\bf L}_{{\bf f}}{\bf -Lb}_\infty
{\bf ,Lb}_\infty {\bf -W}_{{\bf f}}$, or ${\bf W}_{{\bf f}}{\bf -W}_\infty
$ regions quite reliably.

In the region of linear or near-linear finite cluster growth
(${\bf L}_{{\bf f}}$) the considerable increase of $P_2$, is observed
with $E$ increase with relatively weak $P_1$, $P_3$ and $P_4$, vs. $E$
dependences. Such $P_2$ increase is accompanied by the increase of
$R_{_{\max }}$. The ${\bf L}_f {\bf -Lb}_\infty $ boundaries (for
$\lambda =30,$Fig. \ref{fig4}) or ${\bf L} _{{\bf f}}{\bf -Lb_f}$
boundaries (for $\lambda $ $=\infty ,$Fig. \ref{fig5}) at
$E=E_b(\lambda )$ approximately correspond to the region of maximal
$P_2 $ values.

We can estimate the $R_{_{\max }}(E)$ dependence in the ${\bf L}_{{\bf f}}$
region, as well as the $E=E_b(\lambda )$ boundary between the
${\bf L}_{{\bf f}}$ and ${\bf Lb}_\infty $ regions (at finite $\lambda
$) proceeding from the following electrostatic considerations. The
mainly linear clusters grow at small values of $E$ in the ${\bf
L}_{{\bf f}}$ region. In this region the particles tend to join the
chain cluster at the tips where the repulsive energy $U$ is the
lowest. The particles forming a linear chain of length $L$ , repulse
with the energy $U$ the newcomer at the edge of the chain:

\begin{equation}
U\propto \ln (\lambda )  \label{e4a}
\end{equation}

when $\lambda (\leq L)$ is finite, and

\begin{equation}
U\propto \ln (L)  \label{e4b}
\end{equation}

when $\lambda =\infty .$

The growth of the linear chain terminates, when the cluster growth
termination condition (\ref{e2}) is realised, so that we have for the
maximal radius at $E<E_b(\lambda )$:

\begin{equation}
R_{\max }\propto \exp (E)  \label{e5}
\end{equation}

where, the $E_b(\lambda )$ dependency may be approximated for finite
$\lambda $($\lambda <200$) by

\begin{equation}
E_b(\lambda )\cong 1.2\ln (\lambda ).  \label{e6}
\end{equation}

which corresponds to the boundary line between the ${\bf L}_f$ and
${\bf Lb}_\infty $ regions at the diagram of Fig. \ref{fig2}, and
$E_b(\lambda =\infty )\cong 8.0$ is the boundary line between the
${\bf L}_{{\bf f}}$ and ${\bf Lb}_{{\bf f}}$ regions.

Note, that at any $\lambda $ value the Eq.(\ref{e5}) approximates quite well
the behaviour of $R_{_{\max }}(E)$ in the interval of $E<E_b(\lambda )$ (see
Figs. \ref{fig4}b-\ref{fig5}b).

The transition to the region ${\bf W}_f$, where the growth of finite
warm-like clusters takes place (both for $\lambda =30$ and for $\lambda
=\infty $), is accompanied by the sharp decrease of $R_{_{\max }}$ with the
simultaneous increase of the fraction of particles in type (3)
configurations, $P_3$. We approximately localise the
${\bf W}_\infty {\bf -DB}_\infty $ (for $\lambda $ $=30$)
or ${\bf W}_{{\bf f}}-{\bf DB}_{{\bf f}}$
{\bf \ (}for $\lambda $ $=$ $\infty $) boundaries for the transitions
between the warm-like and DBM structures at those values of $E$, which
correspond to approximately the same fractions of particles in
configurations of (2), (3) and (4) types
($P_2\approx P_3\approx P_4\approx 0.2\div 0.3$). Note, that in all
the cases the maximum point of $P_1$ is localised near these values of
$E$. When $E$ is sufficiently large, the type (4) configurations
prevail in the transition zone between the ${\bf W}_\infty $ and
${\bf DB}_\infty $ regions.

\subsection{Radial structure of cluster}

\label{S33}The mechanisms of the internal cluster structure formation, as
well as the matters dealing with its morphology may be clarified by the
analysis of certain cluster properties changes of in the radial direction.
For this purpose, we have analysed the local density profiles $\rho (r)$ and
the $E/U(r)$ ratio that characterise both the cluster growth stability and
the possibility of cluster growth termination if the condition (\ref{e2}) is
realised.

We define the local density profile as

\begin{equation}
\rho (r)=N(r)/N_c(r),  \label{e7}
\end{equation}

where $N(r)$ is the number of cluster particles inside a ring of radius $r$
and one lattice unit width, centered at a seed particle; $N_c(r)$ is the
similar number of densely packed cluster particles. Such definition implies
that the denser is the cluster, the closer is its density to 1.

Consequently, according to such definition, the value of $\rho $ corresponds
to the relative density profile reduced to a density profile of densely
packed cluster. This definition allowed us to suppress the long-range
effects of magic numbers, that are clearly present in the density profiles
of densely packed clusters on regular lattices \cite{smirnov2}.

Figures \ref{fig6}-\ref{fig8} present the profiles of local density,
$\rho $ (a) and $E/U$ (b) for clusters obtained at $\lambda =30,100$
and $\infty $, respectively. These dependencies were obtained at
$E=10,20,...,100$ for each $\lambda $ and averaged over 10 clusters
grown on $200$x$200$ lattice (for $ \lambda =30,100$) and $150$x$150$
lattice (for $\lambda =\infty $). Thus, we have analysed the structure
of clusters in the following regions:
${\bf W}_{ {\bf \infty }}{\bf,DB}_\infty $ and
${\bf C}_\infty $ (for $\lambda =30$),
$ {\bf W}_\infty $ and ${\bf DB}_\infty $ (for $\lambda =100$), as
well as the regions ${\bf W}_{{\bf f}}$ and ${\bf DB}_{{\bf f}}$ (for
$\lambda =\infty $ ) (see the phase diagram in Fig. \ref{fig2}).

In all the cases we have observed a rather sharp and self-correlated initial
decrease of $\rho $ and $E/U$ values with $r$ increase. For $\lambda =30$
and $\lambda =100$ these dependencies have minimum at $r\approx \lambda /2$
and maximum at $r\approx \lambda $. Analysis of the $\rho (r)$ and $E/U(r)$
profiles allows us to make some general conclusions dealing with the
character of internal cluster structure inhomogeneity. In the most general
case, four different zones of spatial cluster structure may be discerned
along the radial direction of cluster structure. (Fig.\ref{fig9}):

zone ${\bf A}$ or internal compact nucleus; this nucleus zone has the
density $\rho \approx 1.0$ and its radius may be estimated as $R_n\approx
E/4 $ (See Appendix, Eq.(\ref{eA4}));

zone ${\bf B}$ or near-nucleus boundary shell; in this shell, the density
drops approximately as $\rho \propto 1/r$ or all the cases observed, and
this shell is located in the region of $R_n<r<\lambda /2$; the existence of
this shell is connected with the growth of near-linear branches from the
dense nucleus;

zone ${\bf C}$, or screening-induced compression shell; this shell is
located in the region of $\lambda /2<r<1.2\lambda $ and may be present only
for finite $\lambda $'s; the density maximum here is observed at $r\approx
\lambda $; this shell exists due to the cluster structure compression
effects, resulting from certain screening limitation of electrostatic
repulsion of particles from the most dense part of cluster(its nucleus) at
distances within the above-mentioned range; correspondingly, the maximum of
the $E/U(r)$ curve is observed in this zone;

zone ${\bf D}$, or external zone; this zone is characterised by the gradual
density decrease and, in some cases, the fractal behaviour of density is
observed: $\rho \propto 1/r^{2-D_f}$, where $D_f$ is the fractal
dimensionality of cluster in this zone (See Section \ref{S34}).

For the case of $\lambda =30$ (Fig. \ref{fig6}), the energy, $E$,
increase from 10 to 100 corresponds to the gradual transition
${\bf W}_\infty \rightarrow {\bf DB}_\infty \rightarrow {\bf C_\infty }$ on
the phase diagram. At $E=10$ (the lowest curve), we are at the border
of ${\bf W} _\infty \rightarrow {\bf DB}_\infty $ (for example, see
Fig. \ref{fig1}, cluster 11). The zone ${\bf A}$ is absent in this
case, since the nucleus radius is very small ($R_n\approx E/4=2.5$).
At $20<E<80$, we are in ${\bf DB}_\infty $ region (for example, see
Fig. \ref{fig1}, cluster 12); here, each energy, $E$, increase by 10
results in a nucleus size increase, approximately, by $2.5$. In these
cases, the profiles of $\rho (r)$ and $E/U(r)$ reveal somewhat
nontrivial behaviour, and we can find the zones ${\bf A\div D}$ in
the cluster structure. At $E\geq 80$, we are almost near the border of
${\bf C_\infty }$ region. We observe that $\rho (r)$ and $E/U(r)$
profiles elevate with $E$ increase; these profiles decrease with $r$
without passing through maxima and minima. The dashed lines on Fig.
\ref {fig6} correspond to the case of $E=130$, i.e., when we are far
in the ${\bf C_\infty }$ region, where the clusters are densely
packed or compact. In this last case, $\rho (r)\approx 1.0$ is
observed and the different zones of cluster structure may be discerned
only by the character of $E/U(r)$ decrease.

The similar behaviour of $\rho (r)$ and $E/U(r)$ profiles is observed
also in the case of $\lambda =100$ (Fig. \ref{fig7}). However, some
aspects of the cluster structure are more explicit in this case. Here,
we have the more prominent maxima and minima of $\rho (r)$ and
$E/U(r)$ profiles, and we observe no general elevation of $E/U(r)$
curves with $E$ increase. It may be explained by the fact that such
behaviour is pronounced only at sufficiently large values of $E$ in
the proximity of compact growth region ${\bf C_\infty }$. We had not
reached ${\bf C_\infty }$ region in our simulations: for
$\lambda =100$, it starts only at $E\geq 300$.

At $\lambda =\infty $, the profiles of $\rho (r)$ and $E/U(r)$ (Fig.
\ref{fig8}) do not pass through maxima and minima, but simply drop
almost like $ 1/r$ outside the nucleus zone, ${\bf A}$. Figures
\ref{fig1}, 21-24 contain good examples of such clusters. The dashed
lines (Fig. \ref{fig9}) correspond to the cases when $E=200$(1) and
$300$(2). In these cases, the clusters have the large compact nuclei
with the radii of $R_n\approx 50-75.$

We have also analysed the behaviour of the $E/U(r)$ profiles for
sufficiently small values of $E$ in such regions as
${\bf L}_{{\bf f}}$, $ {\bf Lb}_\infty $ and ${\bf W}_{{\bf f}}$
for $\lambda =30$ (Fig. \ref{fig10} a) and in such regions as
${\bf L}_{{\bf f}}$, ${\bf Lb}_{{\bf f}}$ and $ {\bf W}_{{\bf f}}$
for $\lambda =\infty 0$ (Fig. \ref{fig10}b). Each of these $E/U(r)$
profiles was obtained through data averaging over the 10 different
clusters. In the regions of finite cluster growth, the $E/U(r)$
decreases quickly with $r$ upraise. At certain critical value of
$r=R_{_{\max }}$, we observe $E/U(r)=1$ at all the points of cluster
perimeter and just at this stage the cluster stops to grow. In
transition zone between the finite and infinite growth regions,
$E/U(r)$ takes a certain stationary value greater than 1.

Remarkable is the case of $\lambda =30$ and $E=4.5$ (Fig.
\ref{fig10}a), when we get into the narrow ${\bf Lb} _\infty $ region
on the phase diagram of Fig. \ref{fig2}. We see that after the fast
initial decrease of $E/U(r)$, its abrupt uprise is obrerved at the
certain $r\approx \lambda $, and then the $E/U(r)$ value of does not
change with $r$. This condition corresponds to formation of the
infinite linear clusters or linear clusters with bending.

\subsection{Fractal properties of clusters}

\label{S34}As we have shown in the previous Section \ref{S33} , in the
general case, the clusters obtained within our model do not reveal any
homogeneous properties in their radial direction. Therefore, the fractal
approach is not applicable for analysis of the structure of whole cluster.
Instead, we have analysed the fractal properties of cluster branches in the
zone ${\bf D}$, or its external zone, where we observed the gradual local
density decrease, which may be approximately described by the relation $\rho
\propto 1/r^{2-D_f}$, where $D_f$ is the fractal dimensionality of cluster
in this local zone.

In order to estimate the $D_f$ value in this zone, we apply the sandbox
method with shells of increasing outer radius, excluding the central cluster
part of the radius $R\approx 1.2\lambda $. We analyse the dependency of the
number of particles in the shell, $N$, versus its area, $S$, using the
following relation:

\begin{equation}
N\propto S^{D_f/2}.  \label{e8}
\end{equation}

The results are averaged over 30 clusters for each value of ($E,$
$\lambda $) pair.

Figure \ref{fig11} presents the results of $D_f$ calculations for
$\lambda =10,30,50,100$ versus $E/\lambda $. On $E$ increase we
observed the transition from linear ($D_f=1$) to compact ($D_f=2$)
structures. The detailed analysis shows that this transition occurs
when the system is in the region where the growth of finite
(${\bf W}_f$ ) or of infinite(${\bf W} _{{\bf \infty }}$) warm-like
clusters takes place. Although the fractal dimensionality $D_f$ can be
slightly lower than $2$ in the ${\bf DB}_\infty $ region, however, it
approaches to this value as the system approaches the
$ {\bf C}_\infty $ region, where the clusters are dense and their
fractal dimensionality $D_f$ is strictly equal to 2. For the case when
$\lambda =\infty $ the branches of ${\bf DB}_{{\bf f}}$ clusters do
not display uniform fractal properties, but their fractal
dimensionality tends to 1 when we move in radial direction from the
centre. So, in this case we have $D_f=1$ and this behaviour is easy to
understand in view of the tendency of electrostatic repulsion to
linearise the branches of the cluster.

We think that the smooth character of $D_f$ transition on passing
through the regions ${\bf W}_f$ or ${\bf W}_{{\bf \infty }}$ results
from the stochastic nature of the model we used in our simulations.
Quite different behaviour of fractal dimensionality can be observed
for the similar, yet deterministic in nature model of aggregation
\cite{lebovka}. In this case we observed the sharp transition between
$D_f=1$ for finite clusters and $D_f=2$ for infinite clusters.

\section{Conclusions}

\label{S4}We have introduced the two-parametric electrostatic Eden model
that simulates the aggregation of charged particles. Cluster growth is
controlled by the screening length of electrostatic interactions $\lambda $,
and the energy of particle adhesion to a cluster $E$. The phase diagram in $
\lambda $ versus $E$ co-ordinates reveals the regions of {\it finite} (f)
and of {\it infinite} ($\infty $) cluster growth. The model gives a variety
of patterns with different morphology, including fractals. We do discern the
following morphologic types: linear, linear with bending, warm-like,
dense-branching and compact Eden-like. The structure compactness increases
with the attraction energy $E$ increase, while the strengthening of
electrostatic repulsion on $\lambda $ increase results in linear
morphologies predominance. The cluster structure is not homogeneous in the
radial direction and in the general case four zones with different
morphological properties may be differentiated: internal compact nucleus
zone, near-nucleus boundary shell, screening induced compression shell, and
external shell.

\acknowledgments{Authors thank Dr. N. S. Pivovarova for valuable discussions
of the manuscript  and help with its preparation.} \label{S5}

\appendix{}

\section{Simple electrostatic estimations for the model}

\label{S6}We approximate a compact two-dimensional Eden cluster by
uniformly charged disk of radius $R$. We calculate the energy of
electrostatic interaction between the charge and compact disk-like
cluster of charged particles $U$ by using numerical integration
instead of summation in Eq. ( \ref{e1}) and taking into account the
homogeneous charge density distribution in the disk-like cluster with
$\rho (r)=1$.

The solid line in Fig. \ref{fig12} presents the dependency of the
reduced electrostatic energy, $U^{*}=U/R$, against the reduced
distance between the charge and the cluster centre, $r^{*}=r/R$, for
$\lambda =\infty $. It is obvious that for the particle located at the
cluster centre, when $r=0$, the energy is equal to $U_o=2\pi R$. At
large distances from the cluster centre, when $ r^{*}\rightarrow
\infty,$ we may neglect the spatial charge distribution inside the
cluster and treat the whole charged cluster as a charge $Q=\pi R^2 $
concentrated at the centre of such cluster. In this case:

\begin{equation}
U^{*}(r^{*}\rightarrow \infty )=U_\infty ^{*}=\pi /r^{*}.  \label{eA1}
\end{equation}

The dependency of $U_\infty ^{*}$ on $r^{*}$ is presented on
Fig.\ref{fig12} by the dashed line. We see, that, at least, for
$r^{*}>1.5$, the Eq.(\ref {eA1}) approximates the $U^{*}(r^{*})$
dependency rather precisely.

If the particle is located at the circle-like cluster perimeter, where
$r=R$ $(r^{*}=1)$, the interaction energy expressions are easy to
obtain in the analytical form for the arbitrary $\lambda $ value:

\begin{mathletters}
\begin{eqnarray}
U_p^{*} &=&4(1+\lambda ^{*}\arccos (\lambda ^{*})-\sqrt{1-\lambda ^{*2}})
\text{ for }\lambda ^{*}<1  \label{eA2a} \\
U_p^{*} &=&4\text{ for }\lambda ^{*}\geq 1  \label{eA2b}
\end{eqnarray}

Here, $\lambda ^{*}=\lambda /(2R)$.

The condition $\lambda ^{*}\geq 1$ from Eq.(\ref{eA2b}) is equivalent
to the condition $\lambda \geq 2R$. In this case, the $U_p$ value is
defined by the interactions with all the cluster charges, without any
screening limitations. Thus, for all the $R\leq \lambda /2$, we obtain
from Eq.(\ref {eA2b}):

\end{mathletters}
\begin{equation}
U_p=4R.  \label{eA3}
\end{equation}

The stable growth of compact aggregates terminates when $E\leq U_p$;
therefore, in the specific case of $\lambda =\infty $ we may estimate
the radius of the compact nucleus as

\begin{equation}
R_n\approx E/4,\text{for }\lambda =\infty  \label{eA4}
\end{equation}

At $R\geq R_n$, the loss of compact cluster growth stability takes
place and cluster ramification begins.

It is possible to estimate the $R_n$ values for finite screening
lengths, $ \lambda $'s, by substituting $U_p=E$ into Eq.(\ref{eA2a}).
However, in this case the situation of infinite compact cluster growth
with $R_n=\infty $ is possible for sufficiently large values of
$E\geq E_c$. For the purposes of $ E_c$ estimation, we consider
Eq.(\ref{eA2a}) when $\lambda ^{*}\rightarrow 0$ (or
$R\rightarrow \infty $ when $\lambda $ is a finite) and obtain
$ U_p^{*}=2\pi \lambda ^{*}$ or $U_p=\pi \lambda $.
So, the relation for $ Ec(\lambda)$ may be written as follows:

\begin{equation}
E_c(\lambda )=\pi \lambda .  \label{eA5}
\end{equation}

The Eq.(\ref{eA5}) allows us to estimate the position of the boundary
between the ${\bf C}_\infty $ and ${\bf DB}_\infty $ regions on the
phase diagram presented in Fig.\ref{fig2}.


%
%

\newpage

\begin{figure}[t]
\caption{Examples of cluster patterns at different values of the attraction
energy $E$ and screening length $\lambda $. Numbers of patterns correspond
to the numbered black squares on the phase diagram on Fig.\ref{fig2}. All
patterns are displayed on the same scale $500$x$500$.}
\label{fig1}
\end{figure}

\begin{figure}[t]
\caption{Phase diagram for the cluster growth in co-ordinates of
screening length $\lambda $  versus attraction energy $E$. See the text
for details.}
\label{fig2}
\end{figure}

\begin{figure}[tbp]
\caption{The four types of the nearest neighbour configurations of cluster
growing on a square lattice discerned for the purposes of configuration
analysis.}
\label{fig3}
\end{figure}

\begin{figure}[tbp]
\caption{The fractions of different types of particles configurations,
$P_i,$ $i=1\div 4$ (a), and maximal cluster radius, $R_{\max }$,(b)
versus attraction energy, $E,$ at $\lambda =30$. The value of $E_b$
determines the boundary between the ${\bf L}_{{\bf f}}$ and
${\bf Lb}_\infty$ regions and may be approximated by Eq.(\ref{e6}). The
value of $E_c$ determines the boundary between the ${\bf DB}_\infty $
and ${\bf C}_\infty $ regions and may be approximated by
Eq.(\ref{eA5}). The dashed line corresponds to the $ R_{\max }(E)$
curve in the interval of $E<E_b$ obtained through rms approximation
using Eq.(\ref{e5}).}
\label{fig4}
\end{figure}

\begin{figure}[tbp]
\caption{The fractions of different types of particles configurations,
$P_i,$ $i=1\div 4$ (a), and maximal cluster radius, $R_{\max }$ $,$(b)
versus attraction energy, $E,$ at $\lambda =\infty $. The value of
$E_b(\approx 8)$ determines the boundary between the ${\bf L}_{{\bf f}}$
and ${\bf Lb_f}$ regions. The dashed line corresponds to the
$R_{\max }(E)$ curve in the interval of $E<E_b$ obtained through rms
approximation using Eq.(\ref{e5}).}
\label{fig5}
\end{figure}

\begin{figure}[tbp]
\caption{The radial profiles of local cluster density, $\rho (r),$ and
the energy ratios, $E/U(r)$. The results correspond to the different
values of attraction energy $E=10,20,30,40,50,60,70,80,90,100$ (solid
lines), $E=130$ (dashed line), the arrow shows the direction of $E$
increase. The screening length is $\lambda =30$.}
\label{fig6}
\end{figure}

\begin{figure}[tbp]
\caption{The radial profiles of local cluster density, $\rho (r),$ and
the energy ratios, $E/U(r)$. The results correspond to the different
values of attraction energy $E=10,20,30,40,50,60,70,80,90,100$ (solid
lines), the arrow shows the direction of $E$ increase. The screening
length is $\lambda =100$.}
\label{fig7}
\end{figure}

\begin{figure}[tbp]
\caption{The radial profiles of local cluster density, $\rho (r),$ and
the energy ratios, $E/U(r)$. The results correspond to the different
values of attraction energy $E=10,20,30,40,50,60,70,80,90,100$ (solid
lines), $ E=200(1),300(2)$(dashed lines), the arrow shows the
direction of $E$ increase. The screening length is $\lambda =\infty$.}
\label{fig8}
\end{figure}

\begin{figure}[tbp]
\caption{The scheme of the internal structure of cluster. We display the
example of cluster pattern and its local density profiles $\rho (r)$ in
radial direction. The intervals of localisation of different zones
${\bf A,B,C}$ and ${\bf D}$ are shown, $R_n$ is the compact nucleus radius,
$\lambda $ is the screening length.}
\label{fig9}
\end{figure}

\begin{figure}[tbp]
\caption{The radial profiles of the energy ratios, $E/U(r),$ for the
screening length $\lambda =30$(a), $\infty $(b).}
\label{fig10}
\end{figure}

\begin{figure}[tbp]
\caption{Fractal dimension $D_f$ of cluster branches in the zone ${\bf D}$
(Fig.\ref{fig9}) versus the attraction energy to screening length ratio,
$ E/\lambda $.}
\label{fig11}
\end{figure}

\begin{figure}[tbp]
\caption{The reduced electrostatic interaction energy $U^{*}=U/R$ versus the
reduced distance between the centre of compact disk-like cluster and unit
charge, $r^{*}=r/R$ at the screening length $\lambda =\infty $. Here $R$ is
the radius of the charged disk.}
\label{fig12}
\end{figure}

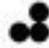
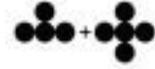
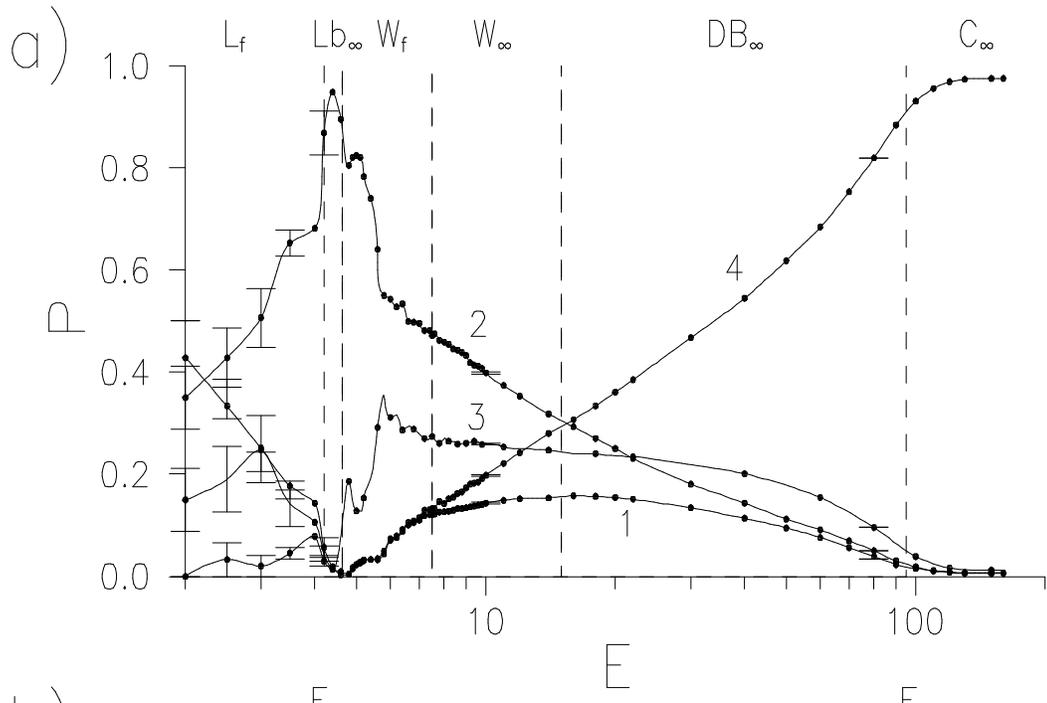
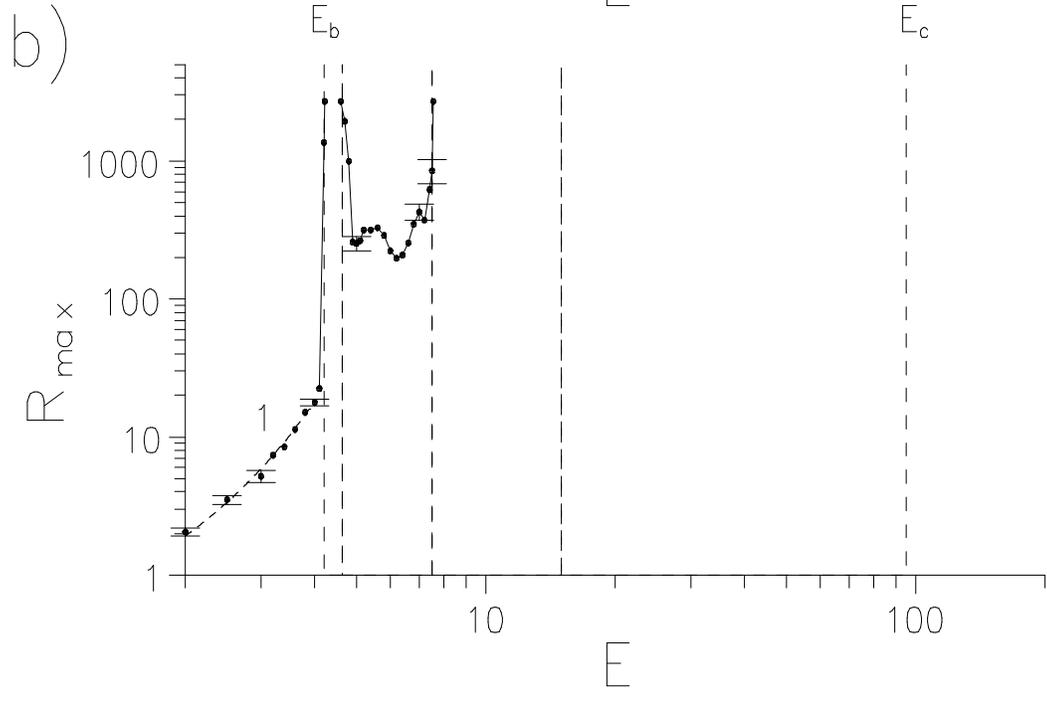
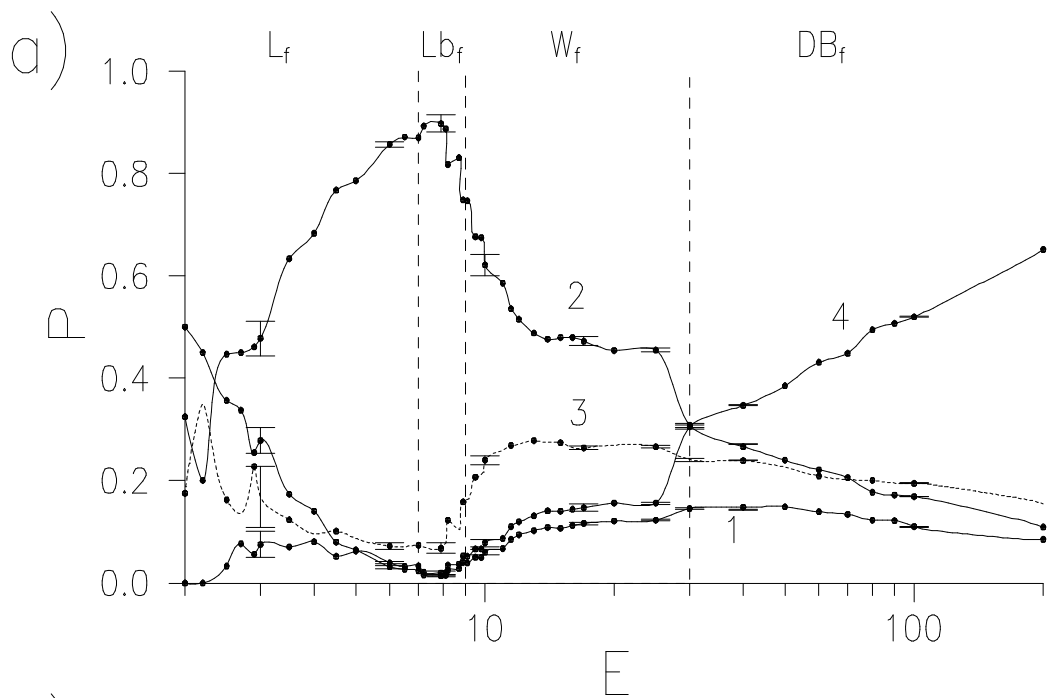
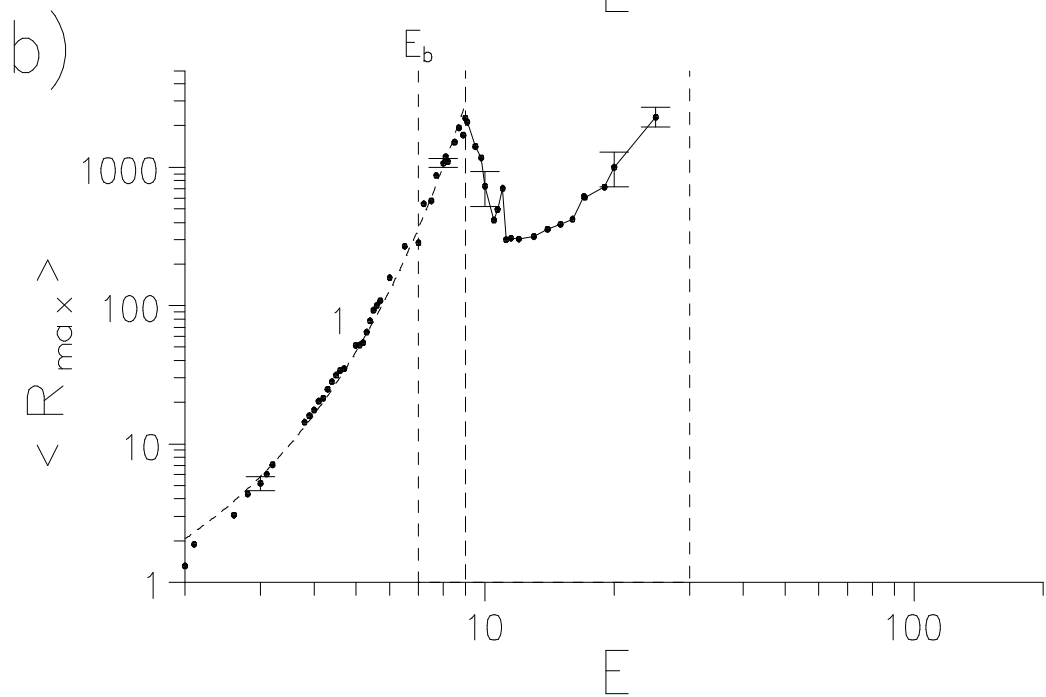
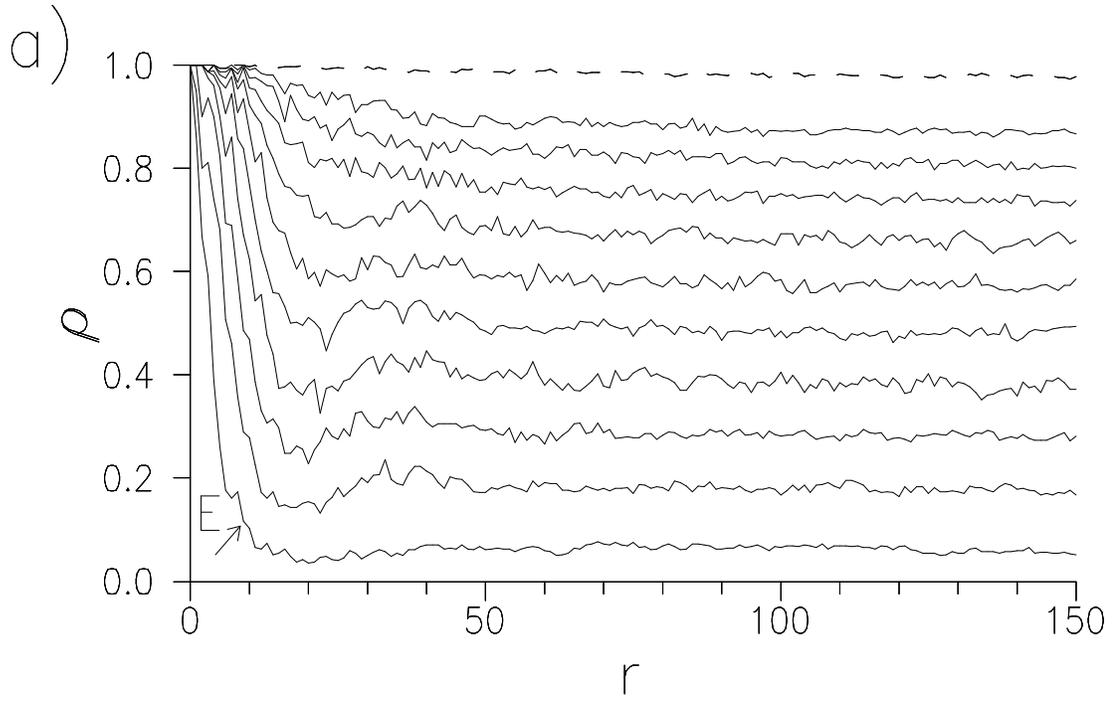
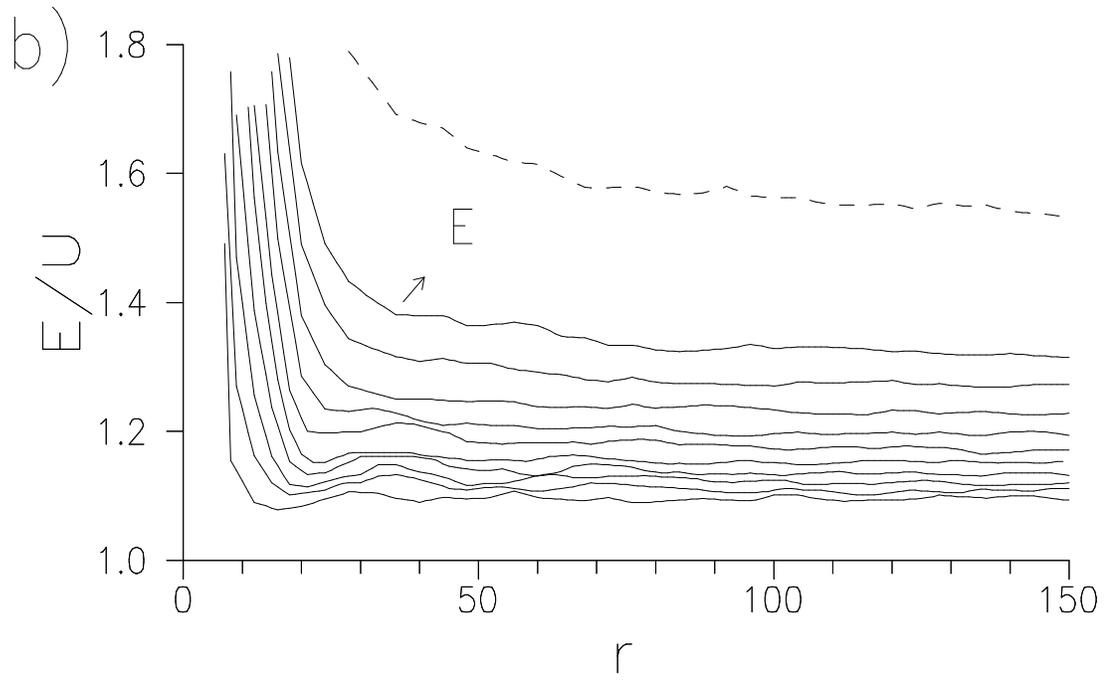
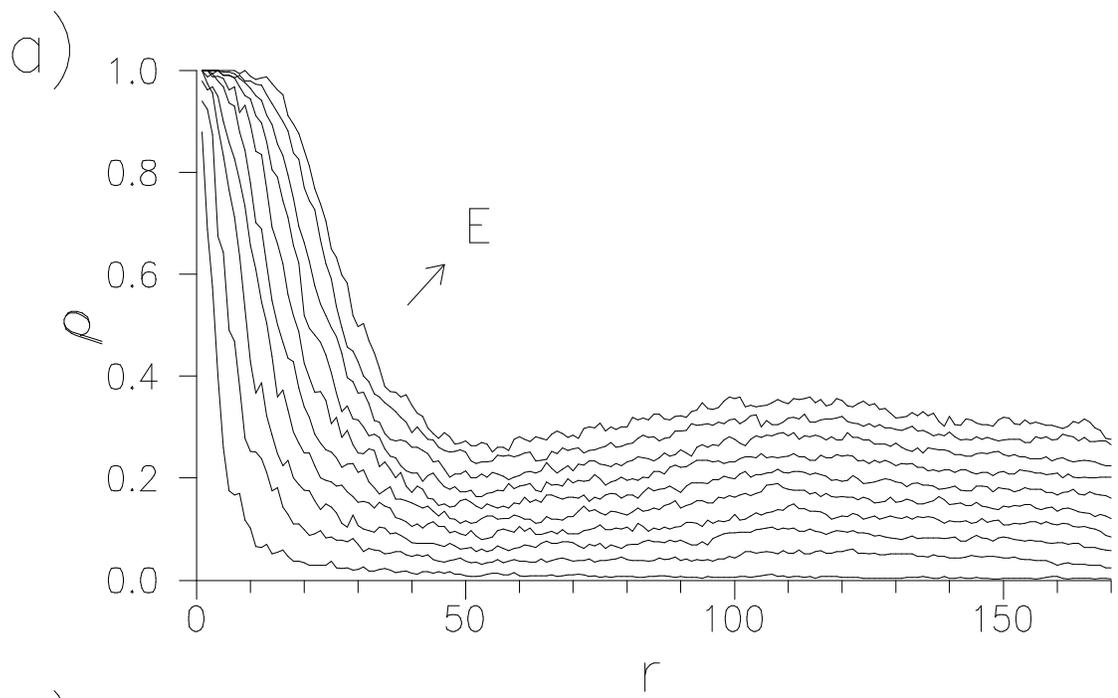
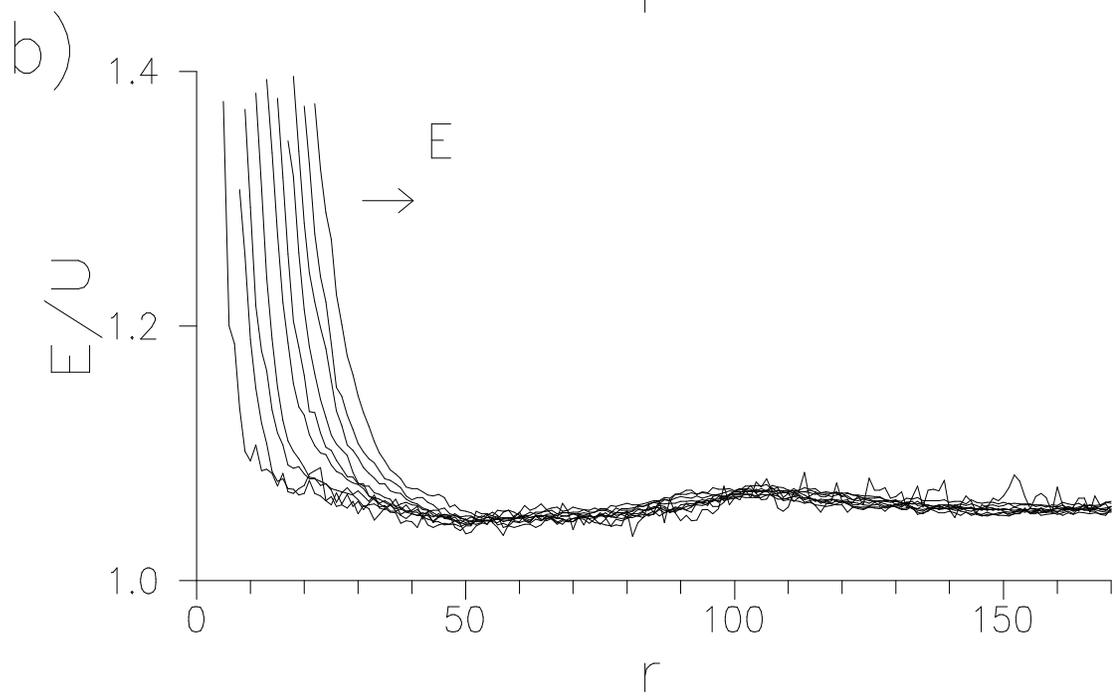
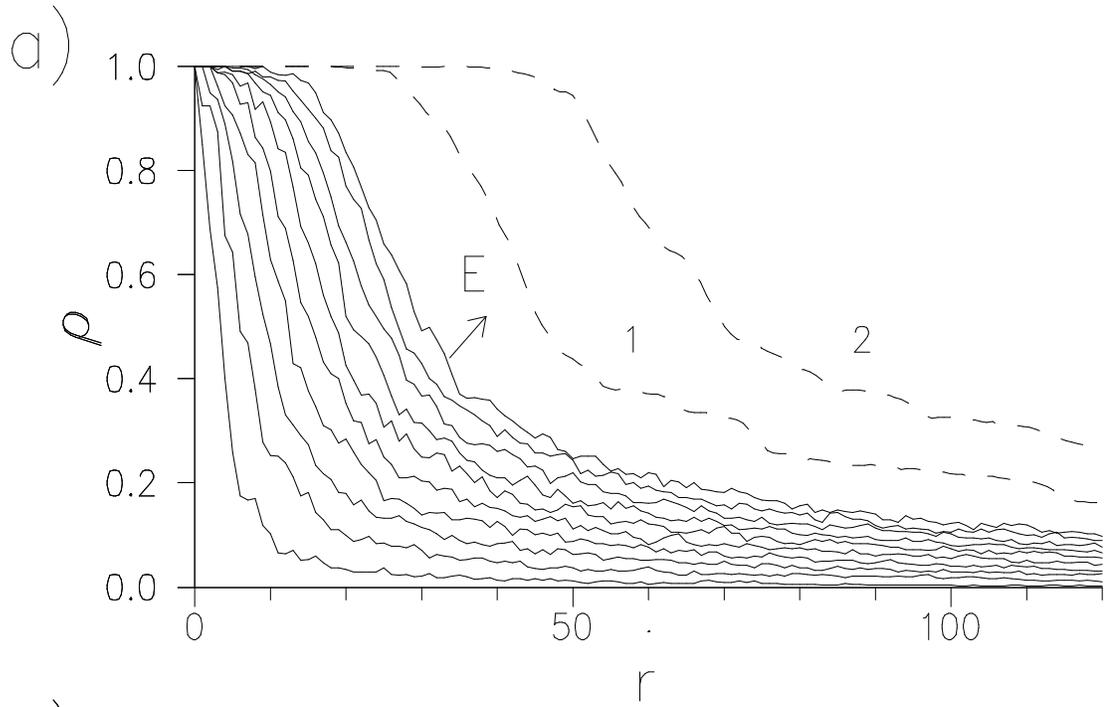
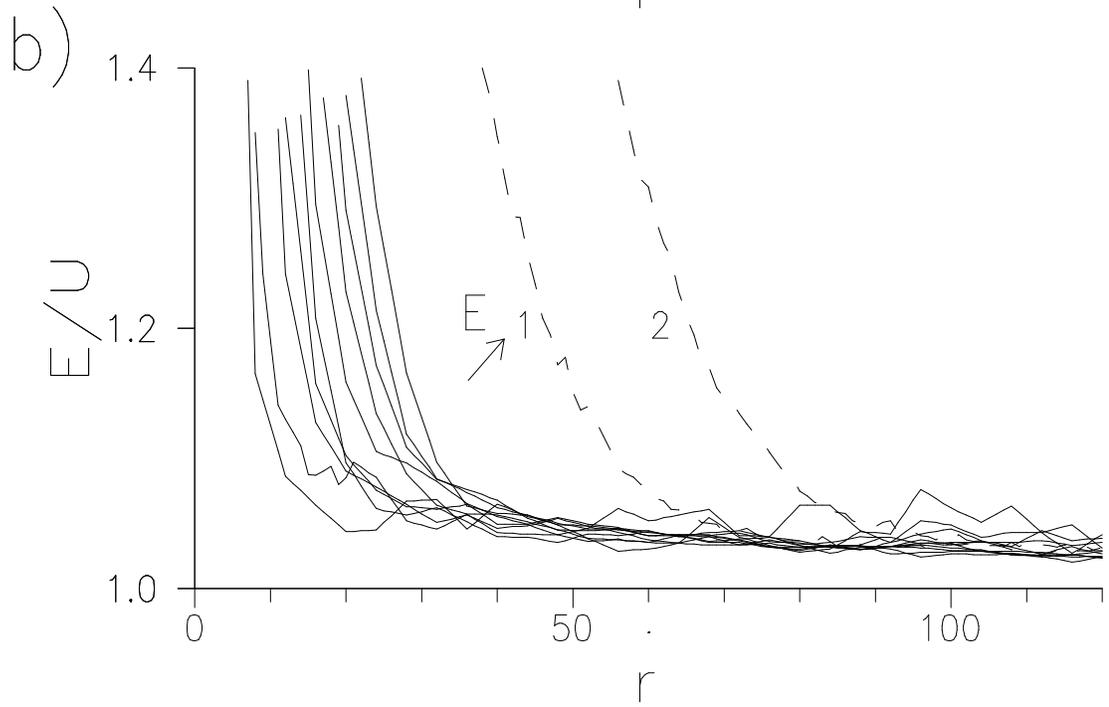
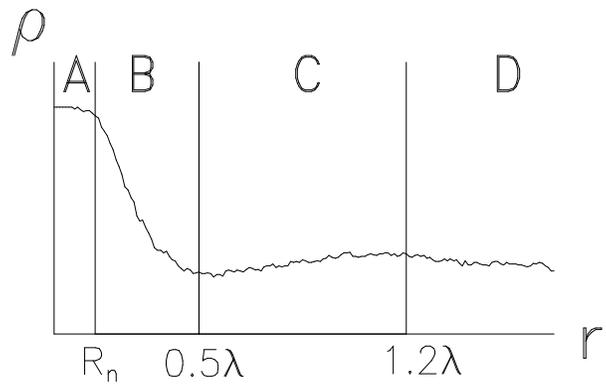
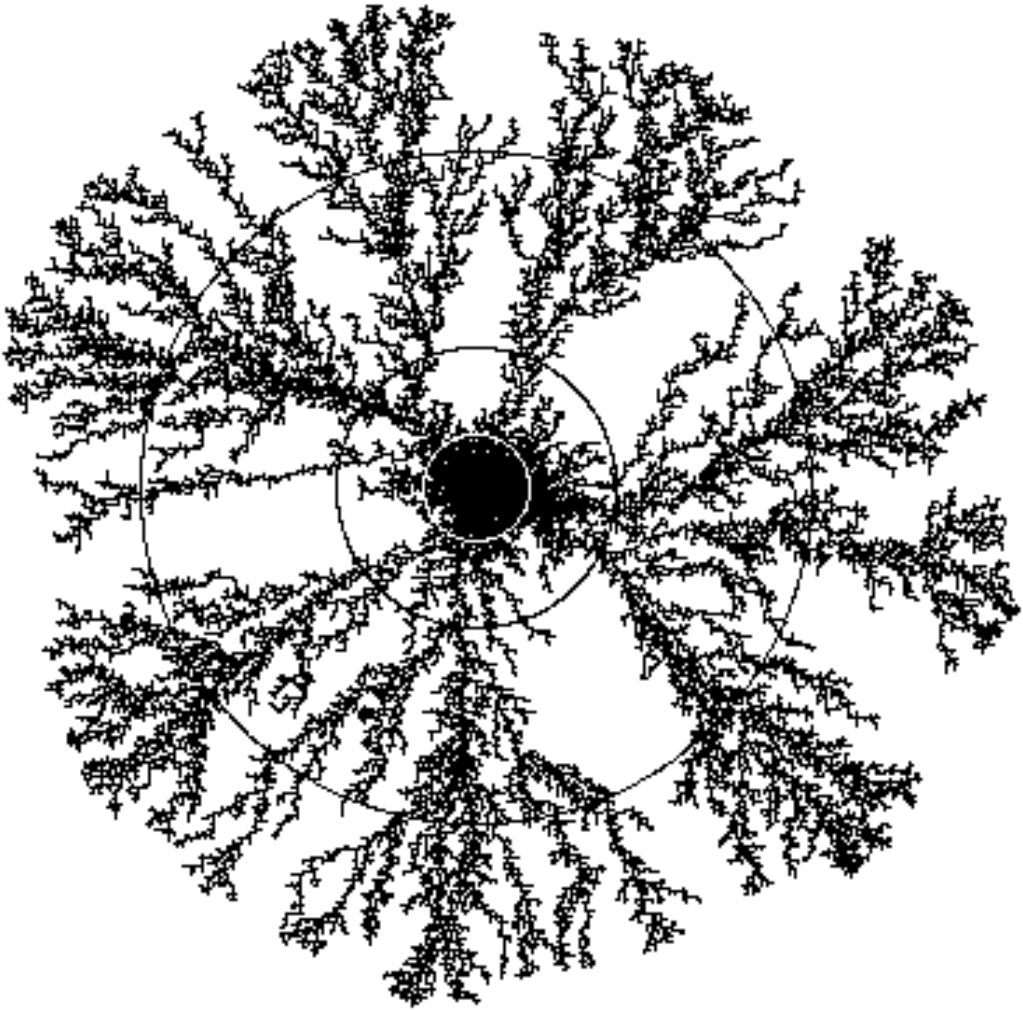
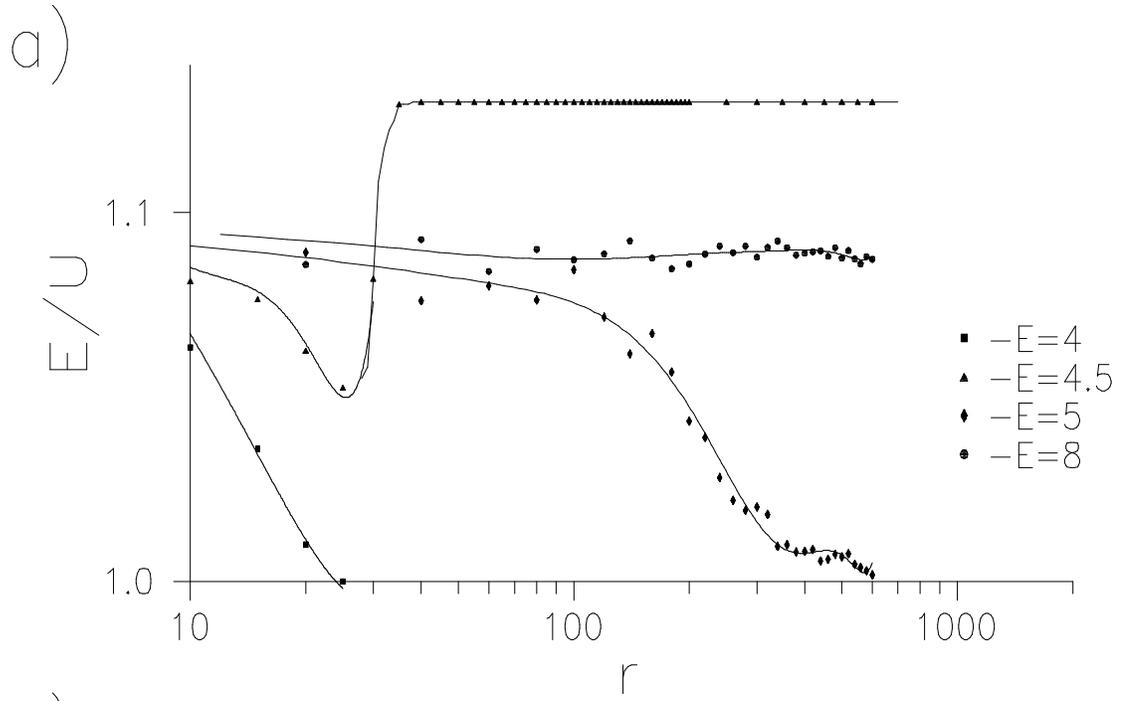
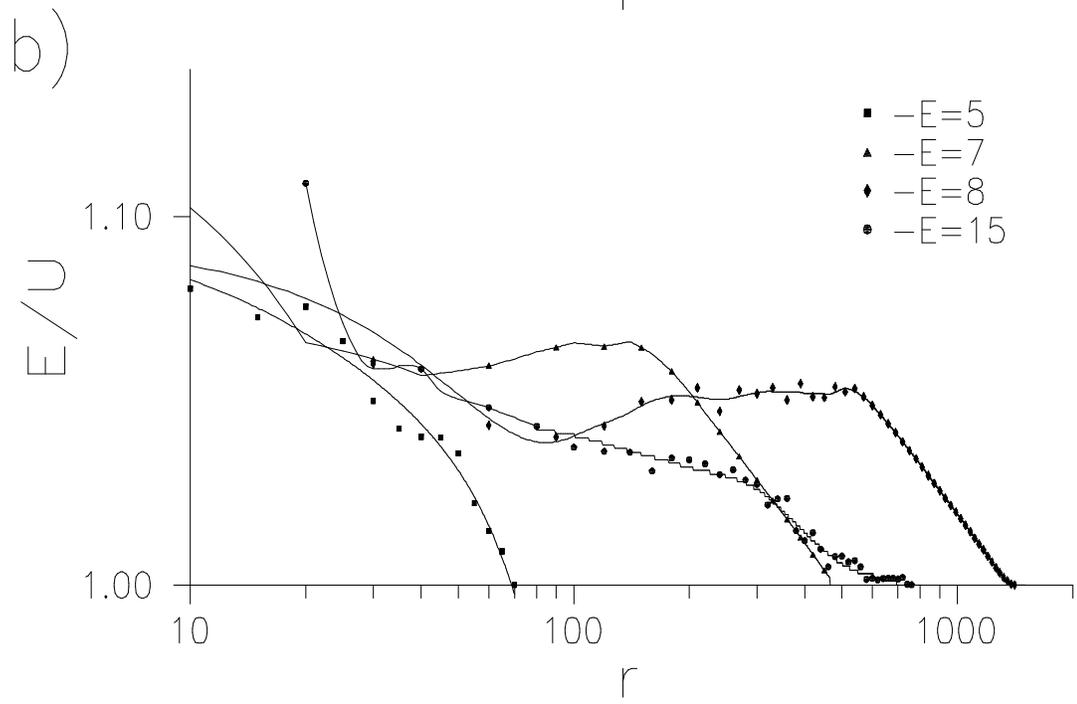
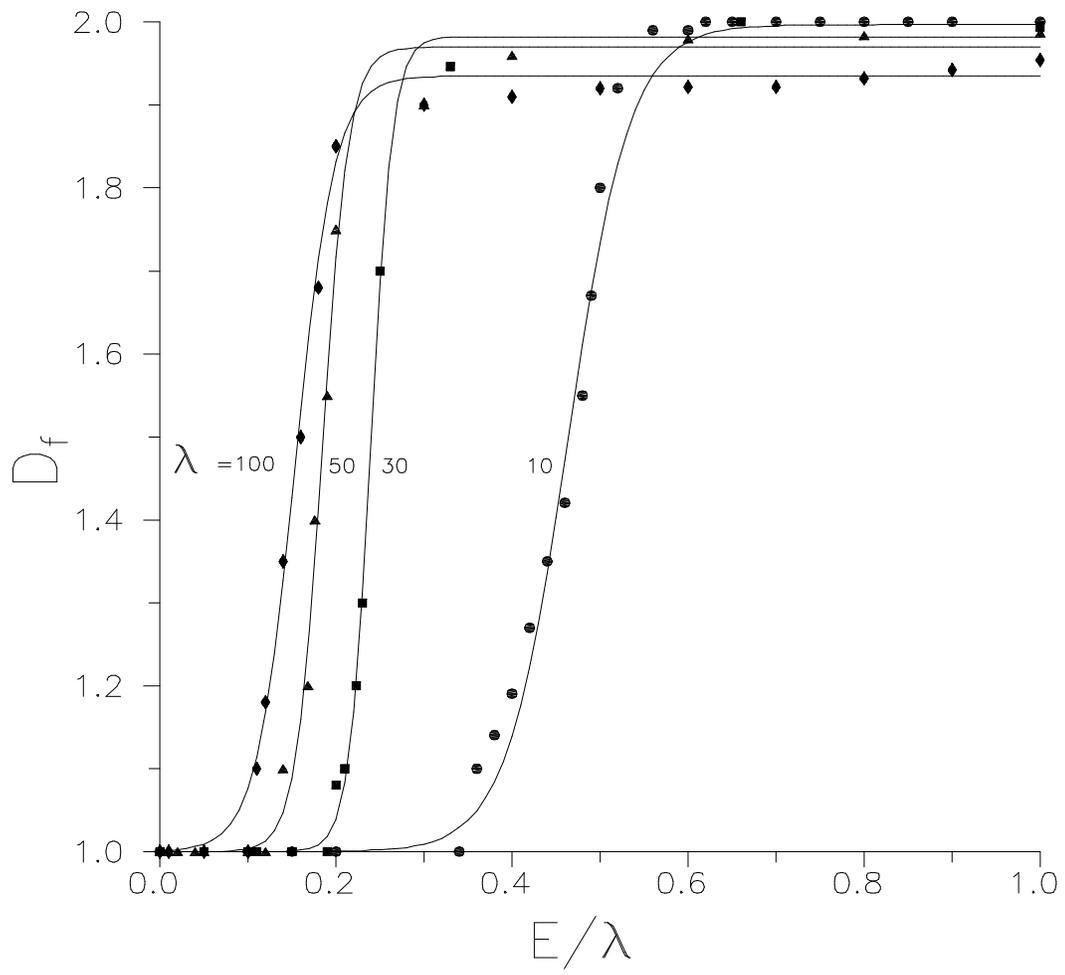
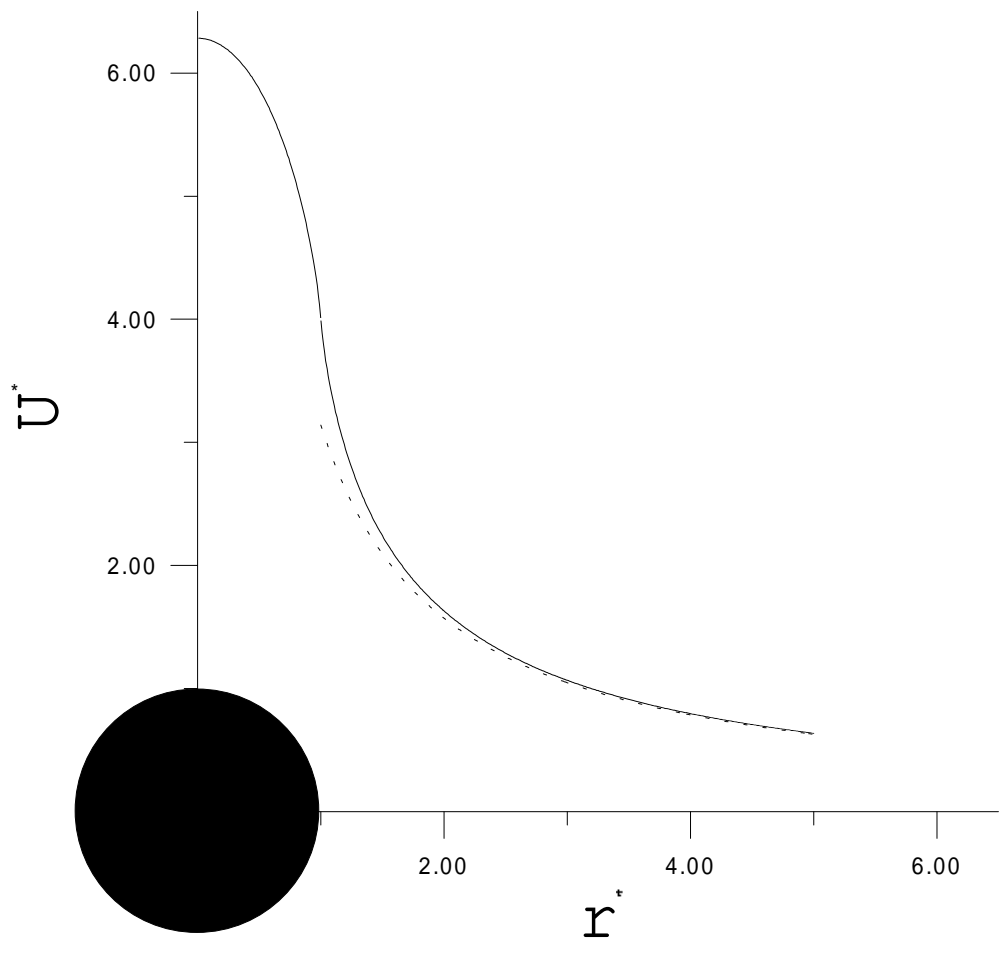
\end{document}